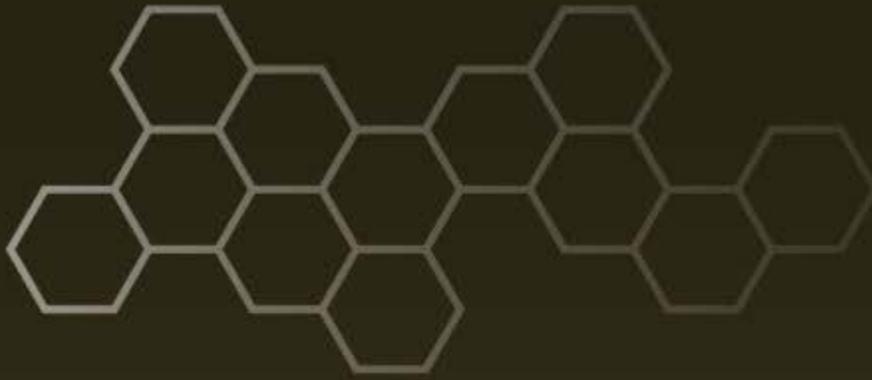
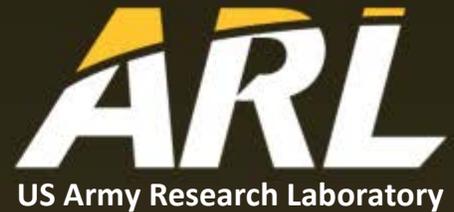

**ARL-SR-0395 ● APR 2018**

**US Army Research Laboratory**

# Toward Intelligent Autonomous Agents for Cyber Defense: Report of the 2017 Workshop by the North Atlantic Treaty Organization (NATO) Research Group IST-152-RTG

by Alexander Kott, Ryan Thomas, Martin Drašar, Markus Kont, Alex Poylisher, Benjamin Blakely, Paul Theron, Nathaniel Evans, Nandi Leslie, Rajdeep Singh, Maria Rigaki, S Jay Yang, Benoit LeBlanc, Paul Losiewicz, Sylvain Hourlier, Misty Blowers, Hugh Harney, Gregory Wehner, Alessandro Guarino, Jana Komárková, and James Rowell

**NOTICES**

**Disclaimers**

The findings in this report are not to be construed as an official Department of the Army position unless so designated by other authorized documents.

The views expressed in this report are those of the authors and not of their employers.

Citation of manufacturer's or trade names does not constitute an official endorsement or approval of the use thereof.

Destroy this report when it is no longer needed. Do not return it to the originator.

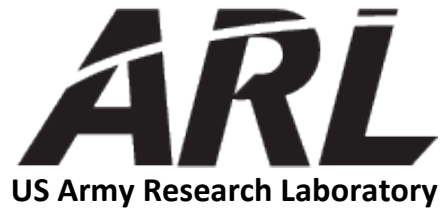

# Toward Intelligent Autonomous Agents for Cyber Defense: Report of the 2017 Workshop by the North Atlantic Treaty Organization (NATO) Research Group IST-152-RTG

by Alexander Kott, *Office of the Director, ARL*

Ryan Thomas, *US Air Force, Washington, DC*

Martin Drašar and Jana Komárková, *Masaryk University, Brno, Czech Republic*

Markus Kont, *NATO Cooperative Cyber Defence Centre of Excellence (CCDCOE), Estonia*

Alex Poylisher, *Vencore Labs, Inc., Basking Ridge, NJ*

Benjamin Blakely and Nathaniel Evans, *Argonne National Laboratory, Lemont, IL*

Paul Theron, *Thales Communications & Security, Gennevilliers, France*

Nandi Leslie, *Raytheon, Waltham, MA*

Rajdeep Singh, *Leidos Commercial Cyber, Foxborough, MA*

Maria Rigaki, *Luleå University of Technology, Luleå, Sweden*

S Jay Yang, *Rochester Institute of Technology, Rochester, NY*

Benoit LeBlanc, *École Nationale Supérieure de Cognitique/Bordeaux INP, Bordeaux, France*

Paul Losiewicz, *DTIC Cybersecurity and Information Systems IAC (CSIAC), Utica, NY*

Sylvain Hourlier, *Thales Avionics and Ecole Nationale Supérieure de Cognitique (ENSC)/ Human Engineering for Aerospace Lab, Bordeaux, France*

Misty Blowers, *ICF International, Fairfax, VA*

Hugh Harney, *Axiom, Inc., Columbia, MD*

Gregory Wehner and James Rowell, *US Naval Research Laboratory, Washington, DC*

Alessandro Guarino, *StudioAG, Cavazzale, Italy*

| REPORT DOCUMENTATION PAGE | | *Form Approved* OMB No. 0704-0188 |
|---|---|---|
| colspan=3 | Public reporting burden for this collection of information is estimated to average 1 hour per response, including the time for reviewing instructions, searching existing data sources, gathering and maintaining the data needed, and completing and reviewing the collection information. Send comments regarding this burden estimate or any other aspect of this collection of information, including suggestions for reducing the burden, to Department of Defense, Washington Headquarters Services, Directorate for Information Operations and Reports (0704-0188), 1215 Jefferson Davis Highway, Suite 1204, Arlington, VA 22202-4302. Respondents should be aware that notwithstanding any other provision of law, no person shall be subject to any penalty for failing to comply with a collection of information if it does not display a currently valid OMB control number. **PLEASE DO NOT RETURN YOUR FORM TO THE ABOVE ADDRESS.** | | |

| 1. REPORT DATE *(DD-MM-YYYY)* | 2. REPORT TYPE | 3. DATES COVERED (From - To) |
|---|---|---|
| April 2018 | Special Report | 18 October 2017–19 March 2018 |

| 4. TITLE AND SUBTITLE | 5a. CONTRACT NUMBER |
|---|---|
| Toward Intelligent Autonomous Agents for Cyber Defense: Report of the 2017 Workshop by the North Atlantic Treaty Organization (NATO) Research Group IST-152-RTG | |
| | 5b. GRANT NUMBER |
| | |
| | 5c. PROGRAM ELEMENT NUMBER |
| | |
| 6. AUTHOR(S) | 5d. PROJECT NUMBER |
| Alexander Kott, Ryan Thomas, Martin Drašar, Markus Kont, Alex Poylisher, Benjamin Blakely, Paul Theron, Nathaniel Evans, Nandi Leslie, Rajdeep Singh, Maria Rigaki, S Jay Yang, Benoit LeBlanc, Paul Losiewicz, Sylvain Hourlier, Misty Blowers, Hugh Harney, Gregory Wehner, Alessandro Guarino, Jana Komárková, and James Rowell | |
| | 5e. TASK NUMBER |
| | |
| | 5f. WORK UNIT NUMBER |
| | |

| 7. PERFORMING ORGANIZATION NAME(S) AND ADDRESS(ES) | 8. PERFORMING ORGANIZATION REPORT NUMBER |
|---|---|
| US Army Research Laboratory<br>Army Research Laboratory (ATTN: RDRL-D)<br>2800 Powder Mill Road, Adelphi, MD 20783-1138 | ARL-SR-0395 |

| 9. SPONSORING/MONITORING AGENCY NAME(S) AND ADDRESS(ES) | 10. SPONSOR/MONITOR'S ACRONYM(S) |
|---|---|
| NATO Science and Technology Organisation<br>Collaboration Support Office (CSO)<br>BP 25, 92201 Neuilly sur Seine, France | NATO |
| | 11. SPONSOR/MONITOR'S REPORT NUMBER(S) |
| | |

**12. DISTRIBUTION/AVAILABILITY STATEMENT**

Approved for public release; distribution is unlimited.

**13. SUPPLEMENTARY NOTES**

**14. ABSTRACT**

This report summarizes the discussions and findings of the Workshop on Intelligent Autonomous Agents for Cyber Defence and Resilience organized by the NATO research group IST-152-RTG. The workshop was held in Prague, Czech Republic, on 18–20 October 2017. There is a growing recognition that future cyber defense should involve extensive use of partially autonomous agents that actively patrol the friendly network, and detect and react to hostile activities rapidly (far faster than human reaction time), before the hostile malware is able to inflict major damage, evade friendly agents, or destroy friendly agents. This requires cyber-defense agents with a significant degree of intelligence, autonomy, self-learning, and adaptability. The report focuses on the following questions:
- In what computing and tactical environments would such an agent operate?
- What data would be available for the agent to observe or ingest?
- What actions would the agent be able to take?
- How would such an agent plan a complex course of actions?
- Would the agent learn from its experiences, and how?
- How would the agent collaborate with humans?
- How can we ensure that the agent will not take undesirable destructive actions?
- Is it possible to help envision such an agent with a simple example?

**15. SUBJECT TERMS**

cybersecurity, cyber defense, autonomous agents, resilience, adversarial intelligence

| 16. SECURITY CLASSIFICATION OF: | | | 17. LIMITATION OF ABSTRACT | 18. NUMBER OF PAGES | 19a. NAME OF RESPONSIBLE PERSON |
|---|---|---|---|---|---|
| a. REPORT | b. ABSTRACT | c. THIS PAGE | | | Alexander Kott |
| | | | UU | 48 | 19b. TELEPHONE NUMBER (Include area code) |
| Unclassified | Unclassified | Unclassified | | | 301-394-1507 |

# Contents

# Acknowledgments

The co-chairs express their gratitude to Ms Katarina Takusova for her outstanding help in organizing and executing the workshop, to Mr John B MacLeod for his critical role in organizing the workshop, and to Carol Johnson for organizing and editing this report.

INTENTIONALLY LEFT BLANK

# 1. Introduction

This report summarizes the discussions and findings of the Workshop on Intelligent Autonomous Agents for Cyber Defence and Resilience organized by the North Atlantic Treaty Organization (NATO) research group IST-152-RTG. The workshop was held in Prague, Czech Republic, on 18–20 October 2017, at the premises of the Czech Technical University in Prague. The workshop was unclassified, releasable to public, and open to representatives of NATO Partnership for Peace (PfP)/Euro-Atlantic Partnership Council (EAPC) nations. The workshop was chaired by program co-chairs Prof Michal Pechoucek, Czech Technical University, Prague, Czech Republic, and Dr Alexander Kott, US Army Research Laboratory, United States.

This workshop explored opportunities in the area of future intelligent autonomous agents in cyber operations. Such agents may potentially serve as fundamental game-changers in the way cyber defense and offense are conducted. Their autonomous reasoning and cyber actions for prevention, detection, and active response to cyber threats may become critical enablers for the field of cybersecurity. Cyber weapons (malware) are rapidly growing in their sophistication and their ability to act autonomously and adapt to specific conditions encountered in a friendly system/network. Current practices of cyber defense against advanced threats continue to be heavily reliant on largely manually driven analysis, detection, and defeat of such malware. There is a growing recognition that future cyber defense should involve extensive use of partially autonomous agents that actively patrol the friendly network, and detect and react to hostile activities rapidly (far faster than human reaction time), before the hostile malware is able to inflict major damage, evade friendly agents, or destroy friendly agents. This requires cyber-defense agents with a significant degree of intelligence, autonomy, self-learning, and adaptability. Autonomy, however, comes with difficult challenges of trust and control by humans.

The workshop investigated how the directions of current and future science and technology may impact and define potential breakthroughs in this field. The presentations and discussions at the workshop produced this report. It focuses on the following questions that the participants of the workshop saw as particularly important:

- In what computing and tactical environments would such an agent operate?
- What data would be available for the agent to observe or ingest?
- What actions would the agent be able to take?

- How would such an agent plan a complex course of actions?
- Would the agent learn from its experiences, and how?
- How would the agent collaborate with humans?
- How can we ensure that the agent will not take undesirable destructive actions?
- Is it possible to help envision such an agent with a simple example?

In addition to this report, the papers presented at the workshop were published as a separate volume, *Intelligent Autonomous Agents for Cyber Defence and Resilience: Proceedings of the NATO IST-152 Workshop, Prague, Czech Republic, 18–20 October 2017*, edited by Alexander Kott and Michal Pechoucek, which can be found online at http://ceur-ws.org/Vol-2057/.

## 2. Environmental Considerations for Autonomous Agent Deployment

Authors: Ryan Thomas and Martin Drašar

With the proliferation of machine-learning (ML) methods in recent years, it is likely that autonomous agents will become commonplace in day-to-day military operations. We expect a significant boost in their capabilities owing to both algorithmic advancements and adoption of purpose-built ML hardware. However, the range of agents' functions will still be, in the foreseeable future, limited by a number of environmental factors, which we attempt to enumerate.

In this section, we recognize 2 types of autonomous agents as 2 extremes on the capability scale. At one extreme are preprogrammed heuristic agents, responding only to specified stimuli based on a set of preset actions. At the other extreme are robust intelligent systems with advanced planning and learning characteristics. Capability is then the aggregate of an agent's intelligence, awareness, connectedness, control, distributedness, level of autonomy, and adaptability.

Environmental factors limit the specific functions and abilities of particular agents and the combination of these factors place an upper bound on agents' capabilities. The following sections provide a list of these factors and their impact.

### 2.1 Mobility

Autonomous agents deployed at stationary structures (e.g., buildings or weapon systems) should suffer the fewest limitations in their operation, as it can reasonably be expected that such agents will have enough power, processing capacity, connectivity, and other resources needed to carry out the most complicated of tasks. These systems will be restricted mostly by the ML state of the art.

Agents deployed on mobile platforms (e.g., vehicles, Soldiers, or missiles) will inevitably be limited by intermittent connectivity; power, space, and processing constraints; or even the physical implications of their actions. Furthermore, for mobile systems, it is likely that the agent will be located at a centralized point in the architecture, rather than be distributed across all subsystems. This is due to the expected difficulty in accrediting systems with robust intelligent behaviors.

## 2.2 Lethality

Agents operating in systems with lethal capacity will either have to undergo much tighter scrutiny or be limited in their actions to prevent the creation of accidental or exploited killer bots. In such systems, it is easy to envision agents and humans performing as a team, with the human having the final authority for decisions with lethal implications. This will require developments in human–machine trust, interfaces, and planning.

Another option to safeguard lethality would be the use of a 2-tier infrastructure, where lethal means are physically separated and thus inaccessible to even a rogue autonomous agent. The ML would control the nonlethal tier only, allowing more conventional means (or, as described previously, a human) to control the lethal tier.

## 2.3 Criticality

Critical systems, whose failure has severe consequences, mostly operate with clear separation of responsibilities and are handled by rigorously trained personnel. Failures are reduced by the application of processes, which limit the impact of human error. Autonomous agents will likely introduce whole new classes of errors, so these error-controlling processes must be updated accordingly.

There are 3 likely approaches to this:

1) Improvements in the understanding of ML operations and performance limits will enable better scrutiny of the inner workings of autonomous agents, constraining the range of possible ML errors and formally proving the scope of exhibited behaviors.

2) Testing methodologies and testbeds will improve, allowing autonomous agents to undergo a battery of conformance tests exhaustive enough to give informal guarantees of the agent's operation with acceptable confidence.

3) Autonomous agents will be deployed redundantly, allowing for robust and resilient operations. Techniques such as voting (e.g., 3 implementations with a majority voting on a next action) could be used.

## 2.4 Connectivity

Most mobile platforms will suffer connectivity problems or forced connection losses. Autonomous agents, which rely on communications links to enable swarm intelligence, command and control (C2), or computation offloading, would be severely impaired during connection loss. Therefore, any such ML functionality requiring connectivity must be designed with respect to the communications environment and timescale in conjunction with required ML decision accuracy. For systems in unreliable environments, which need stable communication channels to arrive at decisions quickly or require accurate and reliable decisions under all conditions, it is up to debate as to whether the presence of autonomous agents is worth the personnel training extension, related updates to operational processes, and associated certification hurdles.

## 2.5 Power and Processing Constraints

Given the currently immense computation requirements for any autonomous and learning behavior, any hardware able to run sufficiently advanced agents will require nontrivial space, power, and cooling. Unless there is a significant leap in technology, this will limit the available resources for agents, especially for deployment in mobile platforms. Developers of agents and policy makers will have to carefully consider which autonomous functionalities are necessary or beneficial enough.

There is great potential in bio-inspired autonomy, assisted by mechanical and structural features on the host platforms. For instance, insects such as moths and flies are an inspiring mix of clever sensor arrays, simple processing cortexes, and advanced mechanical wing design that could enable low-power, low-processing micro-autonomous air platforms.

## 2.6 Commoditization and Standardization of Agents for Environments

We expect that some standard classification of autonomous agents according to their capability and requirements is inevitable. Such classification would ease the adoption process by reducing the need to evaluate each agent in a specific context with regard to whether an agent conforms to a class specification. Military systems then could be limited to specific classes of autonomous agents, thus prescribing the level of autonomy such systems can have.

## 3. Data Sources for Autonomous Agents

Authors: Markus Kont and Alex Poylisher

Possible data sources for autonomous agents can be largely divided into 5 categories: 1) network traffic, 2) event logs, 3) hardware sensor data, 4) operating system (OS)-level software sensor data, and 5) high-level inputs.

Network traffic can be analyzed on several levels of granularity. On the flow level, NetFlow records can be collected from Layer 2 switches and Layer 3 routers (Claise 2004). Where practical in terms of collection/storage overheads, these can be supplemented by full or partial packet captures, and several deep packet inspection methods can then be employed for intrusion detection and traffic indexing.

Event logs can be largely divided between unstructured textual data that would need to be processed with clustering and correlation methods (Vaarandi and Pihelgas 2015; Vaarandi et al. 2015) and structured logs that commonly provide a mix of numerical and categorical data. A subset of categorical event logs is audit log data, whether it be via generic logging of all user commands or tripwires from policy violations, confidential file access, integrity violations, and so on. All major OSs provide this functionality via specialized daemons (e.g., the syslog framework in Linux).

Another major source of information is numerical metrics from hardware performance counters (e.g., central processing unit [CPU] and memory usage, disk and network input/output [I/O], cache hits/misses, and so on) and physical sensors (e.g., temperature, voltage monitoring, GPS location, battery status, and so on). These measurements can also be correlated with measured human responses from sensors attached to human bodies, such as vital signs metrics, which can be quite relevant in a military context, but (like many other aspects) depend heavily on the agent use case (Currie et al. 2017).

OS-level sensor data include system call monitoring, in-memory analysis, kernel execution tracing, protocol state machine tracking for Layer 2 and 3 (Ionescu et al. 2009), and so on. All major OSs provide this functionality (e.g., sysdig, procfs/sysfs, kprobe/jprobe, Berkeley Packet Filter [BPF]/extended BPF [eEBPF]-based tracing in Linux).

Finally, high-level inputs such as mission/goal tracking data are essential in contextualizing other measurements.

It is also possible to implement decoy methods and moving target defense (MTD), even inside a single node, as it is possible to implement this via a simple kernel module that would be able to integrate system call tracking and notification to the higher detection framework (Kerber et al. 2015).

Intrusion detection and sensing tools typically use signatures or behavior analysis (Thompson and Abello 2015; Soudi et al. 2015). Signature-based detection is by far the most commonly used method in practice due to performance considerations, but this method is prone to false positives and the detection quality depends on prior knowledge of attack vectors (Blumbergs et al. 2016). Note that such feeds and analysis methods often feed into each other. For example, network intrusion detection produces categorical event log data that can then be further processed using event correlation and behavior analysis methods. There is also the need for constant aggregation and preprocessing of data. Particularly important are feature extraction, data conversion to a numerical vector space, and the reduction of dimensionality. Datamining and ML methods are simply not designed for processing data in its raw form, thus intelligent and dynamic conversion is needed (Ma and Huang 2017). Exact methods of data collection and processing, however, depend heavily on the use case and require the creation of custom data processing model(s). Note that data collection and sensing modules intertwine heavily with the dynamic learning methods that are to be applied.

The data sources and methods listed previously can be used both for detection and training. However, data acquisition for dynamic learning tasks is a significant and well-known problem. Operational data from real-world systems may not be available due to institutional or legal restrictions. Collecting data from military systems introduces further challenges. Apart from security considerations, the storage space on fielded systems (often embedded) can be very limited. Thus, retention methods may prune viable training data even before the device returns from the mission. Even when available, operational data are almost always unlabeled and incomplete. For example, network connections between compromised workstations and C2 servers are visible from network data, but these connections are often obfuscated and nearly identical to regular background traffic.

Due to the aforementioned difficulties, most academic research in the public domain relies on the relatively small number of publicly available and well-researched data sets. This is a valid methodology for generic datamining and ML research, but it severely impairs the feasibility of using the proposed academic methods on real-world systems or coming up with relevant data models in the first place. As dynamic methods are known to be sensitive to parameter tuning and data set characteristics, a method proven to work on one data set may not work with another (Sommer and Paxson 2010). To make matters worse, label-drift, or classification when features and labels are not known in advance or may change over time, is a significant problem for intrusion detection research and in general for ML in the cybersecurity domain, as in adversarial environments opponents are constantly evolving new attack techniques to bypass defensive methods.

Significant research is currently conducted on dynamic learning and threat detection. However, this work is held back by gaps in validation methods. There is a clear and pressing need to create and maintain up-to-date data sets, which would account for missing entries, the need for correlation, and label-drift problems. One excellent source of such data can be training exercises (e.g., CyberShield), but data collection so far has been secondary to achieving the scenario objectives. Moreover, the timespan of a single exercise is compressed and may not exhibit the long-term temporal patterns needed for anomaly detection (e.g., those of an advanced persistent threat [APT]) and the scope of an exercise is likely to be small in terms of network size.

While actually collecting and labeling more of the exercise data are comparatively easy to address in exercise planning, the scale/timespan/variety of live exercises is by definition limited by resource availability. A complementary approach is to create realistic data sets on demand in emulated/simulated testbeds capable of supporting arbitrary network topologies, sufficiently real hardware, real OS/application/attack/defense software, and a combination of synthetic and real actors. This approach enables a much larger exercise scale/timespan, significant variety, repeatability, and data collection tailored for a particular need by the researchers themselves. A well-known early example of this approach is the Cyber Defense Technology Experimental Research Laboratory (DETERLab; Benzel 2011); a more recent and militarily-relevant example is the Cyber Security Virtual Assured Network (CyberVAN; Chadha et al. 2016), in which multiple diverse labeled and fully reproducible data sets were generated over the past 2 years (Bowen et al. 2016).

## 4. Decision Flow-Based Agent Action Planning

Authors: Benjamin Blakely and Paul Theron

Automated intelligent agents for the purposes of cyber defense may be as varied as the purposes and missions of the systems in which they reside. The determination of an optimal course of action may need to be computed in a small amount of time, with limited resources. An agent may be required to respond to the actions of intelligent actors (human or otherwise) with partial knowledge about the system it is protecting or the impact of its own actions. Here we propose a progressive decision-making approach that takes into account not only the properties of the agent itself, but the environment and context in which a decision must be made. We propose that additional research into the specific criteria for the optimal decision mechanism and acceptance of a proposed course of action, generalizability of such criteria, and determination of how traditional autonomous agent techniques perform in a cyber-defense context, would assist in the development of special-purpose intelligent agents for defending cyber systems from adversarial attack.

The problem of determining the correct course of action for an automated agent has been extensively studied, as summarized, for instance, in Maes (1990) and Ferguson (1995). This research is built on decision theory, including psychological theories of decision making (Edwards 1954), rational choice (Simon 1956; Day 1971), decision making with limited resources in psychology (Kahneman 2002) and computer science (Bratman 1988), and decision making in complex situations when primed by earlier experience (Klein 1988).

From here, the overall construction of agents (Jennings 1998) has been explored. This may focus on their overall behavior, such as logic-based (deduction), reactive (mapping from situation to action), belief-desire-intention (data structures representing these properties of the agent), or layered architectures (software at differing levels of abstraction) (Wooldridge 2013). As we are primarily interested in decision making and planning, it is relevant to look at taxonomies of how agents determine a course of action. Russel and Norvig (2013) separate agents into simple reflex (only using current perception), model-based (environmental models to infer unobservable conditions), goal-based (contributions toward a goal), utility-based (desirability of an action), and learning (improves based on past outcomes) agents.

No single approach is appropriate for cyber defense, and thus an optimal (correctness and efficiency of computation) decision is likely to require progressively increasing difficulty and reliance on mission-specific safety parameters. One possible model would consist of the following stages: Offline Machine Learning, Pattern Recognition, Online Machine Learning, Escalation to a Human Operator, Game-Theoretic Option Search, and Fail Safe. Decisions proposed by each sequential (in our model, but could potentially be extended to parallel, allowing for consensus or value-based approaches) stage are arbitrated for suitability by a decision review stage before being communicated to the relevant actuators (logical or physical). If any stage cannot be executed due to environmental and system parameters or the agent is unable to make a conclusive or acceptable decision, the agent must proceed to the next most-intensive method of decision making.

Loaded into such an agent would be patterns derived from offline learning. This may consist of heuristic or signature-based logic that has achieved sufficient confidence that it allows for making decisions in unambiguous situations. Such offline learning may condense significant amounts of historical and contextual data into a small set of criteria, which can be loaded on potentially low-resource agents. This is an important consideration in scenarios such as network appliances, small drones, munitions, or other systems, which do not have capacity for a large amount of storage and computation onboard. When a set of percepts is to be evaluated by the agent, it will first use this offline learning to determine if the percepts represent a known pattern that can be directly acted upon.

If there is no sufficiently high-confidence match or the proposed action is not acceptable to the decision analysis arbiter, the agent must make a determination as to whether it is in a state conducive to online ML. While this decision is partially static given the parameters of the agent, environmental parameters such as available power, time available to make a decision, or safety margins might make it infeasible to engage a ML engine. If this is not the case, the agent may use ensemble methods such as random forest, or less resource-intensive singular methods, to consider the current state and historical actions and consequences, and make an appropriate decision. The outcome of a decision from any stage, assuming the decision analysis arbiter accepts the decision, would be fed back into this engine so that it can benefit from all actions of the agent.

It may then be necessary to escalate to a human operator to provide alternatives and ask for a decision or indicate that no acceptable options have been determined, requiring human intervention. This is predicated upon either the presence of or path to communication to a human operator, as well as sufficient time for an agent to invoke a human. Otherwise, the agent must skip this step entirely.

If the agent has still failed to reach an acceptable decision, it might yet have enough information (preloaded or from previous stages) to make a reasonable guess as to the outcomes of various actions, even steps ahead. Using this information, the agent can use game-theoretic techniques to construct a short time-horizon decision tree ("short" being relative to the mission parameters and agent resources), evaluate the potential outcomes, and find one that is suitably within compliance, to a given probability, of the desired mission outcome.

Finally, the agent may be required to enter a fail-safe state based upon preprogrammed failure behavior in accordance with mission parameters. Is it acceptable to simply take no action? Take an action with a lower threshold for confidence in efficacy or safety? Terminate, return home, or self-destruct? There is no single action that will satisfy all cases, requiring the terminal stage to be loaded before agent deployment in accordance with the mission profile.

# 5. Autonomous Cyber Agents: Potential Actions

Author: Nathaniel Evans

Agent-based modeling offers a method of embodying a goal-directed cybersecurity expert in an autonomous software program (Macal and North 2005). The software-based agent has goals, performing various actions such as the following to achieve them: evaluating its performance, sensing its environment, performing actions, and communicating and cooperating with other agents and humans. This section describes an autonomous cyber agent with the minor goals of resisting, detecting, and responding to adversaries, all of which work to achieve its major goal of defending and protecting its own system of computers.

To achieve its goals, the agent must incorporate several capabilities. Firstly, the agent needs to perceive its environment by evaluating the state of its computer system(s), detecting changes, and deciding whether those changes are suspicious. If it does indeed detect suspicious changes, the agent must respond in an appropriate and timely manner. Secondly, the agent needs to manage trust relationships with other agents and humans through communication and cooperation. Lastly, throughout its activities, the agent needs to assess itself and its goals, and proactively initiate goal-directed actions to improve its performance and exploit opportunities when possible (Fink et al. 2014).

The following sections describe the 3 main goals of the cybersecurity agent: response, proactive, and social.

## 5.1 Response Goal

The response goal focuses on actions for protection and recovery from a detected suspicious activity. To achieve the response goal, the agent needs to perform a variety of activities, ranging from strict protective measures to tailored increases in protective measures (Yuan et al. 2014). The strictest protective measures include the shutdown of processes or an entire computer system, or disconnecting the computer from communications networks. Other measures preserve several important facets of data security, including confidentiality (through encryption), protection (by scrambling), and availability (by destroying the data). The agent can also restore the system to its last well-known good state.

The agent needs the capability to tailor and increase its protective measures through increased restrictions on configurations, processes, and communication (Bartos and Rehak 2012). For example, changes to configurations can be prohibited, configurations and files can be locked, processes can be isolated, and communications can be restricted by direction (inbound and outbound), location (IP address, hostname, etc.), and characteristics (latency, bandwidth, speed, and amount).

While responding, the agent communicates with other agents to identify and characterize potential adversaries. If possible, the agent should characterize adversaries as benign or malicious, provide a warning of potential foes, recommend a response action, and even receive guidance from other agents on how to respond.

To improve future responses, the agent needs to record and learn from incidents, as well as the actions and outcomes that they instigate. To improve its own resilience, the agent needs the capability to clone itself.

## 5.2  Proactive Goal

The proactive goal lists behaviors that help an agent improve its performance or its protection capability. To achieve the proactive goal, the agent should not simply act in response to its environment; it should be able to exhibit opportunistic, goal-directed behavior and take initiatives when appropriate.

Internally, the agent needs to evaluate itself frequently and apply updates (patches) when necessary.

The agent also needs to assess the state of a system, understand what is normal, and have methods of detecting and characterizing anomalies, including analysis of all event logs. Toward detecting anomalies, the agent needs change-detection tools to discover unauthorized modifications to the filesystem, network device configuration, or application code.

The agent shall assist in the deployment and management of proactive defense technologies such as MTD and cyber deception. Various MTDs can benefit from oversight by autonomous cyber agents, including those that focus on process rotation (e.g., Morphisec), application rotation (e.g., Dynamic Application Rotation Environment [DARE] MTD), OS rotation (e.g., Multiple OS Rotational Environment [MORE] MTD; Thompson et al. 2014), IP rotation (e.g., Morphinator), and network path rotation (e.g., Stream Splitting MTD). Agents can also employ various cyber-deception methods, including direct observation hiding, process hiding, data hiding, network hiding, investigation hiding, and social hiding.

## 5.3  Social Goal

The social goal describes how and why an agent should communicate and cooperate with other agents and humans. This collaboration essentially improves the overall protection of a given computer system (Jennings et al. 2014).

The primary communication ensures the sharing of information that will allow other agents to understand the state of the computer system, its adversaries, and its responses. The shared information includes whom and what has been blocked, what has been allowed, and the associated signatures that have been used to create these blocks and allows.

Other systems need alerts from the agent on the actions that were taken and the expected effects of said actions. The agent needs to maintain awareness about the state of other agents, including uptime and the ability to respond by cloning peers.

Establishing and managing trust relationships enable agents to cooperate and rely upon each other. The agents need methods to manage those trust relationships, including the ability to create, break, and detect violations within trust relationships.

Agents shall interact, when appropriate, with other artificial agents and humans in order to complete their own problem solving and help others with their activities.

## 6. Applying Machine-Learning Algorithms to Cyber-Physical System Security

Authors: Nandi Leslie, Rajdeep Singh, Maria Rigaki, and S Jay Yang

Cybersecurity analysis is transforming to support cyber-physical system (CPS) security: the needs for protecting the data confidentiality, integrity, and availability have accelerated with the increased autonomy and connectivity of engineered systems. Yet, there are significant research gaps in CPS security. Several research questions are what specific ML algorithms exist for securing CPS processes and data, whether considering adversarial learning and cyber deception can enhance CPS security, and to what extent can useful cyber-defense information be learned to predict or prevent cyberattacks. For CPS requiring standalone functionality over extended periods of time, the entities tend to have significant constraints on computational resource use for network, battery power, and memory usage, which also impact the design of the cyber systems monitoring and securing substantive amounts of data. In this case, when considering methods for CPS defense, whether for commercial or mission systems, traditional intrusion detection methods are required yet inadequate for ensuring that these systems operate efficiently and securely without human support or maintenance. For example, in an Internet of Things (IoT) environment, mobile Android devices can function successfully with lightweight intrusion detection systems (IDSs)—these IDSs operate with minimal computational resources by using n-grams of packet payloads to classify malicious network traffic combined with supervised linear classification and Bloom filters (Wang and Stolfo 2004; Chang et al. 2013; Yu and Leslie 2017). Furthermore, replacing signature-based IDS with these computational algorithms can improve IDS performance, including efficiency and prediction performance (Wang and Stolfo 2004; Chang et al. 2013; Yu et al. 2014; Yu and Leslie 2017). It is known that ML and anomaly-based detection algorithms can improve detection accuracy because these computational methods allow for the detection of novel cyberattacks, where the exploits were previously unknown or no signatures exist (i.e., "zero-day" cyberattacks), which is by definition infeasible with signature-based or misuse detection models for IDS (Lunt 1993; Wang and Stolfo 2004). However, these anomaly-based algorithms for IDS tend to have high false-positive rates (Lunt 1993)—benign packets are frequently misclassified as malicious—requiring human analysts to review the IDS alerts in more detail.

The December 2010 report of the President's Council of Advisors on Science and Technology (PCAST) included cybersecurity, ML, and CPS among the research priorities proposed. Furthermore, the PCAST report recommends that US Government agencies, the National Science Foundation (NSF), and the Defense Advanced Research Projects Agency (DARPA) take steps to ensure data privacy and confidentiality for CPSs by developing cybersecurity models using statistical/ML and computer science that are characterized, in part, by what adversary actions can be withstood. In this section, we explore 3 capabilities for software agents (Genesereth and Ketchpel 1994) that can advance CPS security in adversarial environments with limited network connectivity: 1) autonomous cybersecurity provided by software designed to monitor network security and perform other cyber-defensive actions, which resides at the host and operates with intelligence and flexibility (Balasubramaniyan et al. 1998); 2) cyber-risk quantification, which is lacking in many operational systems (Leslie et al. 2017); and 3) adversarial learning to address the fact that cyber systems are often reactive, excluding adaptive adversarial decision making and actions that can specifically influence intrusion detection and prevention. We refer to this cyber system for CPS as an "autonomous cyber-defense agent (ACDA)" throughout.

## 6.1 Autonomy

To advance autonomy, ACDAs for a CPS network with limited interconnectivity should automatically switch between processes (e.g., intrusion prediction, IDS, and antivirus), learn and adapt from their local environment with real-time data, and share estimates and predictions asynchronously. When connectivity is available, these agents should use online learning or batch-training techniques; whereas during periods of isolation, agents would benefit from operating with fewer features in the testing set than those available during training. In a mobile ad hoc network (MANET) with Android devices, for example, it is known that using this batch-training concept, where agents are downloading and sharing information with their peers and C2, can improve efficiency (Chang et al. 2013; Yu and Leslie 2017). ACDAs should be able to defend against novel or anomalous cyberattacks for which IDS/intrusion protection system (IPS) signatures do not exist, for instance, by using anomaly-based detection and unsupervised and semi-supervised learning models (Balasubramaniyan et al. 1998). Probabilistic graph models (PGMs) provide scalable solutions for estimation in coordinated environments to handle issues like rumor propagation (Kschischang et al. 2001), which may ensure cybersecurity. However, additional research is needed to validate a successful application of PGMs to cybersecurity and risk assessments.

Although an ACDA should be able to detect cyberattacks with network- and host-based systems combined with signature- and anomaly-based detection models, cyber situational understanding and awareness can enhance the autonomy of these entities by integrating cyber-risk quantification to improve ACDA predictions regarding the cyber threats and vulnerabilities along with adversarial learning algorithms—these computational models simulate the adaptive decision making between agents and their adversaries—that can influence the full interconnected process for cyber-risk quantification.

### 6.2 Cyber-Risk Quantification

Enhancing an ACDA with intrusion prediction and forecasting models can transform the computational algorithms from being solely detection focused to incorporating statistical learning models to determine which network activities are correlated with and predictive of cyberattacks, including attack graphs (i.e., Bayesian), variable length Markov models, semi-supervised learning, and generalized linear models (Strapp and Yang 2014; Yang et al. 2014; Leslie et al. 2017). Modeling cyber vulnerabilities can involve predicting the number of successful cyberattacks and the lateral movement of malicious traffic through a network. Furthermore, it is known that predicting the lateral movement of an attack within the CPS can provide specific details about the cyberattack timing and methods. Figure 1 shows the steps involved in cyber-risk modeling.

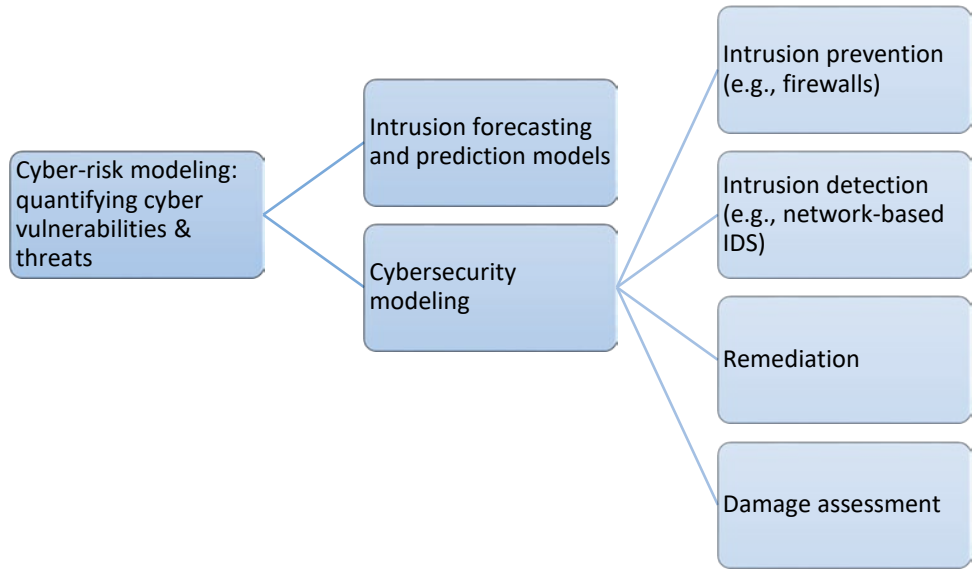

Fig. 1   Cyber-risk modeling

Measuring and examining the similarities between successful cyberattacks—this includes attack sequences, tactics, and overarching adversarial strategies—is an additional ACDA capability that can reduce exposures to cyber risks. Specifically, advancing ACDA cyber-risk assessments involves understanding how well observed attack sequence can be "explained" based on other attack sequences (e.g., Kullback–Leibler [K-L] divergence). For example, the DARPA Explainable Artificial Intelligence (XAI) program (Gunning 2016) focused on improving the ability of ML models to explain black box decision making to end-users providing the level of trust needed in operating with these autonomous systems (Lei et al. 2016). This includes developing techniques, such as Local Interpretable Model-Agnostic Explanations (LIME; Ribeiro et al. 2016) to learn more structured, interpretable, and causal models as well as discriminative ML models to infer and interrogate explainable models from a black box (Guo et al. 2017).

More recently, the Intelligence Advanced Research Projects Activity's (IARPA's) Cyber-attack Automated Unconventional Sensor Environment (CAUSE) program (Rahmer 2015) calls for forecasting of cyber incidents using unconventional signals. The performers have preliminary results to demonstrate the extent data such as public sentiment, social media trends, news reports, open threat exchange update, dark web transactions, and so on can help forecast future cyber incidents. For example, through the use of reconfigurable Bayesian networks with an innovative way to deal with imbalanced data, Okutan et al. (2017) show that approximately 70% area under curve (AUC) can be achieved to forecast the occurrence of specific cyberattack types on specific organizations. In addition, Werner et al. (2017) show the use of time-series analysis using autoregressive integrated moving average (ARIMA) can be effective in some limited cases.

### 6.3 Adversarial Learning

Learning in the presence of adversaries implies that ACDAs need the ability to defend themselves against cyberattacks during both training (i.e., data and model poisoning) and deployment with adversarial samples. In addition to classifying the attackers in terms of their knowledge of the learning models (Papernot et al. 2017) or access to the data used for learning (Barreno et al. 2006), we can also define the capabilities of an ACDA in terms of robustness to deception based on prior knowledge of adversaries' capabilities (McEneaney et al. 2007, 2008).

While actual attribution of malware authorship or attacks can be a difficult task due to varying levels of attacker deception tactics, learning about their characteristics can be a more useful approach. Source code, call graph, and domain similarity, as well as other types of semantics, can be used for that purpose. Learning about adversarial tactics and behaviors can be used to predict an adversary's next moves. Empirical adversarial models can be generated using semi-supervised learning (Strapp and Yang 2014) and likely scenarios can be produced using simulation of adversarial behaviors (Moskal et al. 2017).

In addition, robustness to deception requires that agents address unknown-unknowns, and to do so, agents must simultaneously address scalability. For example, for IoT, millions of devices each requiring the security provided by an ACDA could benefit from curse-of-dimensionality-free methods—these computational algorithms provide scalability by transforming the information state and functions over the information states into idempotent algebras and using highly efficient pruning techniques (McEneaney 2009). Furthermore, resource-constrained agent networks need to address deception with a distributed deception robust control approach (Lawson et al. 2015), starting with smaller/multiple local games spanning a smaller state space using a priori knowledge and expanding them dynamically with observations (Horák et al. 2017). In many cases, cyber robustness is a more feasible goal than intrusion detection, and game-theoretic modeling provides mission-centric robust decision making in the face of deception using ML for improving the information state spanning (McEneaney et al. 2007). Similarly, beyond just defending and detecting adversarial attacks, an ACDA must be able to adaptively taking measures such as launching of cyber deception as a defensive and preventative action (Durkota et al. 2016). For example, agents can have value-based tasking and use game-theoretic and optimization approaches to ensure that the information gain/value is maximized with respect to the goal of the agent (McEneaney et al. 2008).

## 6.4 Conclusion

Increasing numbers of CPS and IoT devices must operate securely in contested environments with limited connectivity. To address these cybersecurity and risk challenges, our proposed ACDA for each host has 3 main capabilities that are designed to advance CPS security: autonomy, cyber-risk modeling (Strapp and Yang 2014; Leslie et al. 2017), and adversarial learning (Moskal et al. 2017; Papernot et al. 2017). Cyber-risk and security modeling enhance the situational understanding for the Internet of Battlefield Things (IoBT)—this process involves using computational algorithms to predict and detect successful network intrusions—by quantifying the combined impact of cyber vulnerabilities and threats on a networked and engineered system. Moreover, the framework that we propose here for an ACDA also uses adversarial learning and game-theoretic modeling techniques to incorporate advanced adversarial decision making and cyber deception and advance cyber-risk and security quantification. These capabilities are critical for assessing how to increase autonomy in CPS/IoT defense and may help ensure that connected entities are interoperable, reliable, and secure.

## 7. A Program for Effective and Secure Operations by Autonomous Agents and Human Operators in Communications-Constrained Tactical Environments

Authors: Benoit LeBlanc, Paul Losiewicz, and Sylvain Hourlier

### 7.1 Background: The Problem Description

We consider the following question: What specific means can help human operators deal with the intersection of remote interactions with autonomous agents in bandwidth- or emission-constrained tactical communications environments, with the goal of satisfying cybersecurity and emissions control (EMCON) requirements?

### 7.2 Plan for Solution

We employed a 7-step process to design an applied research and development program to address the problem:

1) Clearly identify the applied autonomy problem and goals.

2) Propose relevant formalisms with which to describe the problem.

3) Provide a tactical context within which to bound the problem.

4) Describe a method for cognitive resource management of the tactical communications constraints.

5) Propose a knowledge representation scheme.

6) Co-evolve development of the specific autonomous agent platforms and the cognitively informed human–machine interface (HMI).

7) Design a training program for both autonomous agents and human operators. This should be iterative in the course of the development of the agents and the HMI.

## 7.3 Details of the Solution

### 7.3.1 Problem Description

Effective use of intelligent agent autonomy is the proposed means to offset communications constraints imposed by tactical considerations. However, this autonomous capability needs to satisfy the necessary and sufficient conditions for adequate situational awareness, security, and battle management by both local and remote human operators. It is proposed that autonomy be applied adaptively to the tactical situation at hand, using an extension of a cognitive resource management approach developed by the Ecole Nationale Supérieure de Cognitique (ENSC) and Thales.

### 7.3.2 Relevant Formalisms

We proposed 2 mature candidates: Unified Modeling Language (UML) by the Object Management Group (2015) or Entity Relationship Models (ERM) developed by Peter Chen (1975), now at Carnegie Mellon University.

### 7.3.3 Tactical Context

We proposed the use of predefined tactics, techniques, and procedures (TTPs) for cybersecurity and operational security (OPSEC) community human operators (e.g., cyber operations for cyber mission forces [CMFs]). We have also employed an exemplar tactical context by using an exemplar system for cockpit interactions based on work carried out at ENSC.

## 7.4 Proposal of a Cognitive Resource Management Model (CRMM) of HMI with Autonomous Agents

We sketch out a CRMM evaluated to be both adequate and fruitful for use by both autonomous agents and human agent operators (Fig. 2). This CRMM was the most controversial aspect of our 7-step process. It necessitated a model that was relatively mature, but *analogical*, insofar as the cognitive primitives and processes of both software agents and human agents could only be analogically mapped into each other at the lower levels of the model. We proposed an extension of one of the cognitive models previously developed by Sylvain Hourlier of the Human Engineering for Aeronautics Laboratory (HEAL; Hourlier and Suhir 2014). This work was jointly developed by ENSC and Thales. It was evaluated to be appropriate for use in this domain as 1) it had already been implemented and tested for HMI use by Thales Avionics in the AV2020 cockpit demonstrator, and 2) it was felt to be sufficiently intuitive to provide a fruitful analogical model for application to autonomous agent HMI.

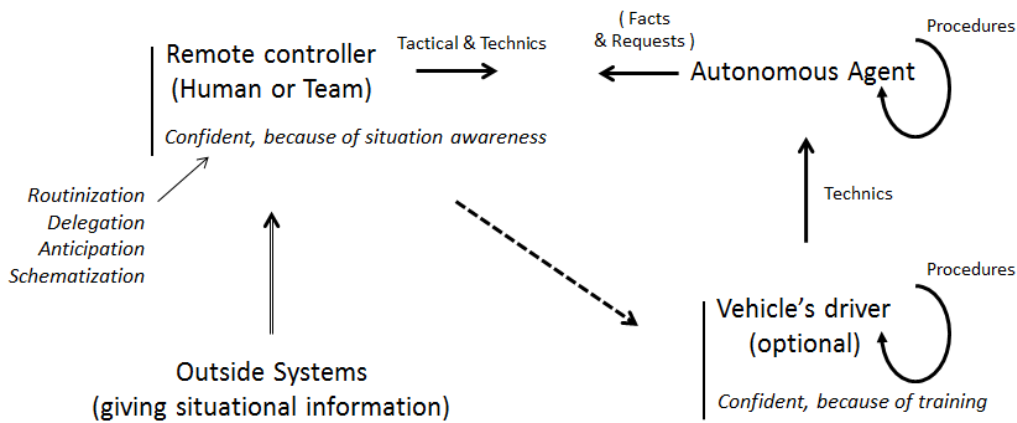

Fig. 2   CRMM process

In the CRMM, the 4 classical human strategies that favor an economical use of operator resources have been analyzed and transposed for this application: anticipation, routinization, schematization, and delegation. The original application was to incorporate them into the design of avionics to facilitate their use, hence favoring better pilot cognitive resource management. The ambition here is to analyze these 4 strategies and provide for their opportune application to human–autonomous agent collaboration. As the CRMM was developed initially at an individual level, it is a logical step up to extend the model to a collective multiple-agent collaboration.

The goal of the CRMM is to point to the efforts that we must do in systems development to preserve and optimize the cognitive resources of users. Doing that, the human user is adequately prepared to commit their own cognitive resources to the realization and consequences of the agent's action within the context of a larger plan.

The following describes an exemplar hierarchical CRMM:

1) The base resource level is composed of reflex routines (i.e., purely autonomous reflexive procedures, carried out by an agent in communications-constrained environments, with possibly no reporting by the agent or acknowledgement by the human operator based on EMCON). They are presumed to be nonoverrideable by the human operator. Having reflexes or automatic routines enables agent autonomy. Full knowledge of these reflex actions improves the human user activity, transforming them, by extension, into an actor in tactical role playing. Full situational awareness by the human operator may be constrained by the EMCON state, however.

2) The second resource level is composed of semi-autonomous subroutines potentially integrated into different specific plans, with planner oversight. These are predetermined routines but with oversight (i.e., they are not reflexive actions; the agent needs to be told to execute them from among alternative actions). The agent needs to receive either a monodirectional order to execute, or if not available, an a priori default procedure "previsioned" and decided by its own local sensor input. The human operator would be aware of what the previsioned defaults are, and which, in the appropriate communications environment, could be overridden by the human operator. Anticipation is an opportunity for the operator to put their own personal indicators in a kind of timeline, which could be developed during the action. The human operator is primed to be ready to interoperate.

3) The third resource level incorporates situation-specific collaborative schemas or decision trees with critical decision points that would require *collaborative input* for the agent to execute (i.e., the agent has to wait for appropriate information to execute, as the agent cannot execute based on its internal sensor input alone). In these cases, the agent requires offboard data for a specific situation, and no a priori or default action is possible, such as using satellite missile targeting data for over-the-horizon (OTH) targeting. Schematization is a way to install pretested procedures related to a tactical action, involving diverse reactions, with minimal communications exchange.

4) The highest level envisioned is that of delegation. The agent is here reporting aspects of its internal situational awareness of its own capability. The agent then delegates an action to be performed by another agent or human operator (e.g., the agent is unable to carry out a predefined subroutine or collaborate in a situation-specific schema, and delegates the response to an alternate). Delegation adds a kind of autonomous service support, activated as needed, within larger actions. The human operator is expecting to be offered the lateral use of alternative agents as part of their operational toolset.

All of these levels are inherently emissions-correlated action plans (i.e., the level of autonomy and human operator interaction is correlated to the EMCON levels in effect).

## 7.5  Context-Informed Knowledge Representation

An adequate knowledge representation scheme for HMI in communications-constrained tactical environments was evaluated to be the most mission- and platform-specific area within this research program. Our strongest insight was that we needed compact representations of the information required to satisfy operational constraints for individual autonomous platforms and human operators within their C2 architectures. These necessary and sufficient minimal representations could be designed a priori by referring to mission- or platform-specific TTPs, but would need validation for adequacy by high-fidelity modeling and simulation (M&S) and actual training exercises.

## 7.6  Co-Evolution of Agent Autonomy and the HMI

The recommendation is for iterative design and development of agent autonomy along with the HMI to support it. This includes modeling, simulation, and prototype testing long before systems are sent to operational test and evaluation (OT&E).

## 7.7  Modeling and Simulation and Training

The requirement for effective M&S and operational training with autonomous systems goes without saying. It should be acknowledged by systems developers that there may be unanticipated emergent effects that may arise with regard to human interaction with autonomous agents. Effective M&S of just autonomous agent behavior may be easier to carry out, as we can constrain their behavior when we build them. The interaction of autonomous agents and human operators is less reliably modeled, due to the possibility of emergent behaviors found in complex adaptive systems (CASs) (Holland 2006). This is due to the adaptive behavior of all the elements in the system, including the human operators.

## 8. Preventing the "Killer Robot" Scenario

Authors: Misty Blowers, Hugh Harney, and Gregory Wehner

What specific means could prevent an autonomous agent from evolving into a killer bot or exercising some other undesirable behavior? While the notion of "killer robots" may be dramatized by science fiction and popular culture, the potential for autonomous systems to produce fatal results is a very serious concern. From emerging technologies, such as self-driving vehicles, to established examples, such as shipboard automated missile defense systems, autonomous devices with the capacity for causing unintended harm are a reality.

Autonomous agents hold the promise of highly efficient execution of complex tasks, with the implication that the agents can perform the task as well as or better than a human counterpart. The other side of that promise is that these agents will behave in ways we cannot predict. For any autonomous agent scenario, the risk of harm can be determined as a factor of the reach of impact of the system affected by the autonomous agent. Impact is the measure of unintended harmful consequences of the affected systems. For example, for an autonomous weapons system, impact is the damage the weapon can cause and reach is the range of the weapon. Reach must not exceed the boundary of need—which is determined by the value of the system mission—and need must be considered against the consequences of potential impact.

Consider an air-gapped autonomous agent deployed on a remote device in a dangerous environment. The agent has the capability to administer the remote device in the absence of human oversight based upon mission parameters and the data it collects from its environment. The risk may be assessed by quantifying and characterizing the potential impact of the remote device the agent administrates. Is it a weapon system, a communications relay, or a radar station? Can the components be manipulated by misuse or neglect to produce dangerous conditions? Do those conditions outweigh the need met by the system? In the example of an air-gapped autonomous system, there is a perception that the "reach", or ability to affect other systems or assets, is limited. However, this is not always the case. The "Stuxnet" worm is a perfect example of how a cyber payload designed for a specific target system made its way to the outside world and had unintended effects.

Another scenario is a "bot", or self-contained software program in a distributed network, capable of making decisions on what actions to take to either defend or attack a network environment. It may base these decisions upon information that it senses and learns from the network environment or from information it receives from other "bots" operating in the same environment. Such agents are often driven by a type of utility function, such as minimizing the number of nodes in this network environment that are compromised by malware. For a networked device, risk is harder to define—as the range of potential impact to consider may extend to every device in the network. What are the implications if the "bot" spreads to the wide-area network? Depending on the capacity of the agent, software boundaries may limit the reach, but it is prudent to be cautious of this assumption. Risk in this scenario is tightly bound to the level of effect the agent can produce on the network and other networks that are connected to it. Consideration should also be given to a peer-to-peer, or robot-to-robot, oversight, which may result in more sophisticated solution spaces than those constrained by human comprehension. The distribution and networking of these autonomous devices extend the basic logic, allowing group interactions that can operate counter to the mission as stated.

We therefore frame prevention as the reduction of risk of impact and the resilience to both intentional and unintentional compromise. An unintentional compromise is the result of a malfunction or unforeseen circumstance for which the system is ill pre-prepared. An intentional compromise is the result of a malicious act that results in an effect outside of the intended design or purpose. Prevention can be broken into 2 main approaches: restriction, or bounded autonomy, and self-guidance driven by some ethical code of behavior.

Some members of the scientific community believe that autonomous capabilities should be designed with boundaries and limitations, while others believe that the learning mechanism should not be constrained until sufficient M&S has been conducted to allow the autonomy to discover complex solution spaces that would have been outside the bounds of human definition. As an approach to prevention, restriction is clearly the first design decision that must be taken into account.

Limiting autonomy produces a fundamental reduction of impact, yet it may not allow the system to meet the mission need. Other restrictions may applied for unshackled agents—mechanical governors on device range, resource restrictions that limit impact, or network restrictions to limit mobility. Supervision is a form of external restriction, where manual overrides, backdoors, and input dependencies for execution tie impact to human judgment. The most important restriction may be on market deployment. The promise of benefits cannot be allowed to outstrip the rigors of research and development. Fielding devices with agents whose capacity for impact is not fully understood would be a failure of humanity, not artificial intelligence.

The notion of self-guidance approaches the field of robot ethics. How can autonomous agents be imparted with decision rulesets that guide their own evolution and interactions, and how can those rulesets be partitioned off from the range of effect. It is hard enough to determine ethical guidance regarding human agents, and even with clear rules, humans do not always follow them. Let's assume we somehow design an autonomous agent to follow the Geneva Convention—what happens if the device sensors that identify humans malfunction or are deliberately sabotaged? What if an autonomous car is deceived into identifying sidewalks as roads? Can an autonomous agent reach the conclusion that they are causing impact and assess their own guidance protocols? If those protocols have been corrupted, what action should the agent take? It may be the case that these rulesets be partitioned in such a way that any change to them immediately causes the agent destroy itself, and perhaps even the device it affects.

Looking ahead, there seem to be 2 types of these peer-to-peer interactions that need to be understood. The first behavior is related to operations as a cooperative system rather than as an isolated device. This system-level behavior can be understood if we examine network gateways that we will assume are autonomous. Each gateway may learn that a reduction in performance may be the appropriate response to sudden increases in traffic from a single source. However, if all gateways have this logic, then the response to an emergency state may result in the gateways reducing performance exactly at the time it is needed the most. Basic research is needed to capture human-like judgment ontology and logic. This is a challenging topic given that judgment logic needs to be embedded on independent devices all operating with imperfect knowledge. At first blush, a structure for heterogeneous logic is required to facilitate distributed mission accomplishment. This mirrors the judgment of humans; people apply heterogeneous logic and thereby avoid some fragile vulnerability states.

The second behavior relates to the autonomous system interacting with the embedding environment. Again, the autonomous system needs to develop some human-like characteristic that is arguably rare in even human systems—judgment of reasonable appropriate response. Consider a distributed net of military robot guardians protecting a resource. An adversary probes the area. The guardians have the option to not respond (preserving the secrecy of location), respond with minimal force, or respond at total capability. The judgment normally performed by humans will consider the situation at hand, then make a value judgment. Basic research is needed to develop a definition, creation, and testing of this judgment logic. Advances in game theory point to some attractive research that could lead to advances in this logic. Choosing appropriate objective and reward functions may also be critical to the evolution of behaviors. To add to the complexity, this logic further needs to account for distributed autonomous processing and execution.

If we look at independent agents operating in a system using similar logic and requiring similar resources, we then could consider a classic CAS. CASs have been studied using very simple logic. There are classic behaviors that seem to occur across observed systems (from agriculture to epidemiology). These systems tend to create organized behaviors that work while the system has ample resources; however, these systems tend to show changes in organization under stress, leading to catastrophic failure modality.

Current testing and simulation models need basic advances to facilitate multilevel tests. Today, atomic device testing would validate proper functionality in isolation. Research is needed to define a peer-to-peer networked functionality definition and compositions adequate for validation of autonomous group functionality in multiple complex scenarios. Finally, the autonomous group must be tested in realistic emulations of the larger environment where it is intended to operate, not just to assess its risks as one agent, but the risks that arise from a single agent's interactions with a larger distributed system. This will require an ontology spanning multiple environments with sufficient complexity to define success and failure conditions. More test and evaluation are necessary to reduce risk of the "killer robot scenario" and allow the community to develop autonomous capabilities that can incorporate complex human judgment with compassion and vision.

**Further Reading**

## 9. "Hello, World" Autonomous Agent

Authors: Alessandro Guarino, Jana Komárková, and James Rowell

The challenge we tackle in this section is the design of an actual autonomous agent, small and simple to implement but able to illustrate the essential functions any autonomous intelligent agent (AIA) should possess, albeit in a streamlined way. The agent proposed here is a purely software agent intended for cyber defense only.

To be a proper AIA, it should fulfill the following 6 characteristics:

1) An agent is strictly associated with its environment: an autonomous agent outside the environment it was designed for can be useless, or not even an agent at all. Franklin and Graesser (1996) have given a convincing definition of agents and the ways in which they differ from other software. The first 4 points in our definition draw from their definition.

2) An agent interacts with the environment, via appropriate sensors providing input from it and appropriate actuators, allowing the agent to act and influence that environment.

3) An autonomous agent acts toward a goal, or, in other words, it has an "agenda". In particular, an autonomous agent developed for warfare operations is assigned a target.

4) The activities of a truly autonomous agent are sustained "over time", so it must have a continuity of action.

5) An autonomous agent should possess an adequate internal model of its environment, including its goal—expressed possibly in terms of world states—together with some kind of performance measure or utility function that expresses its preferences.

6) An agent must possess the capability to learn new knowledge and the possibility to modify over time its model of the world and possibly also its goals and preferences.

In this section, we describe the agent and explain how it fulfills these requirements. We define its environment, task, and properties, such as sensors and actions. We also discuss possible extensions of the agent.

To make these "Hello, world" autonomous agents feasible, the design makes specific assumptions about the environment in which the agent operates, and the number and type of inputs and outputs its sensors and actuators will have. This has the aim of keeping the complexity low.

## 9.1 Environment

AGENTX lives in a virtualized cloud environment that supplies some unspecified cloud-based service. We assume this platform runs 3 kinds of virtual machines (VMs, or virtual servers): database servers, application servers, and web servers. We also assume that a hypervisor exists to manage the platform and balance the load.

## 9.2 Task

Again, for the sake of simplicity, AGENTX performs one specific function and not in an open-ended generic network defense mission. Its goal is to manage a set of honeypot (HP) virtual servers with the objective to deceive adversaries and deflect cyberattacks against the cloud platform. Its architecture is monolithic (as opposed to a distributed, swarm-like structure) and operates at the hypervisor level of the system. To perform some of the available actions, AGENTX relies on small applets installed on each virtual server, for instance, exposing a RESTful application programming interface (API). It must be noted that in the context of this proof of concept, security measures that in a real environment would be mandatory are overlooked (e.g., encryption of communications, self-protection of the agent itself, and so on).

Since the mission of AGENTX is purely deception, it implements the capability of communicating to other autonomous agents (and/or to human supervisors) the necessity to intervene and implement active defense measures.

The agent has access to background information, such as a set of ready-made HP images, dummy process containers, and dummy files.

## 9.3 Sensors

The sensors are able to access the following data and information:

- Alerts from IDSs (count and severity)
- Integrity information of critical files on the VMs
- Metadata about critical files on the VMs

- Processes
- Log files
- Metrics on the level of use of resources and system load
- Feedback and replies from other agents tasked with active measures

## 9.4 Actions

The following actions are available to the agent:

- Starting and stopping HP VMs.
- Starting and stopping actual virtual server instances (optionally).
- Initiating a "cry for help" message to other agents (or humans).
- Deploying dummy files and applications, and quarantine files (via the applets).

## 9.5 Learning

The agent implements a reinforcement learning model employing an appropriate reward function:

$$R = a\frac{honey\_events}{security\_events} + b\frac{\Delta\_resources}{total\_resources} + c\frac{justified\_CFH}{CW}, \quad (1)$$

where

*honey_events*: metric for attacks/events against the HPs;

*security_events*: number of attacks against the real servers (detected by IDSs);

*total_resources*: metric for the total amount of resources available;

Δ*_resources*: resources freed or needed (example, spinning HPs) to implement an action by the agent (negative when resources are needed, positive when resources are freed);

*justified_CFH*: "justified cries for help", number of messages (alerts) sent by the agent reacting to actual attacks; and

*CW*: "cry wolf", number of messages sent by the agent requesting assistance for attacks that did not really happen.

The coefficients *a, b, c* state the relative importance of each factor. They should be tuned beforehand or during the initial learning phase.

We consider total resources available as those actually available at the time of action, which makes the function and the agent's behavior dynamic during that time. It also means that a relatively costly action is not penalized if the system is under very light load, because the number of available resources is high and even small actions are heavily penalized if the system is utilizing almost all its resources.

Note that this function could be calculated—in a future version of AGENTX—for homogeneous groups of VMs (e.g., only the application servers), to better reflect the situation and the world state, providing AGENTX with a more granular and detailed view of its environment.

The learning is performed by implementing an anomaly detector leveraging a small set of hard-coded features (for the purposes of this section) including the following:

- Number and severity of IDS alerts
- Anti-malware software alerts
- Unauthorized accesses
- Access to HPs or dummy files
- Alerts from dummy processes
- File integrity violations
- System load (aggregate, by group, and individual)

## 9.6 Testing

To validate the performance of the agent, we have to set up a testing environment, perform real attacks, and evaluate its efficiency. Since the agent is learning with each attack, we should let the evaluation continue for some time so the learning process can take place. It would also be ideal to face the agent with real attackers, not only simulated attacks.

We propose to validate the agent on defending a network with several servers in a virtualized environment with simulated "regular" traffic. The setup has the following advantage: since we know which traffic was generated by us, we can safely assume the rest of the traffic comes from the attacker; therefore, we can easily recognize the justified and unjustified cries for help. The detection part is also easily achieved in this setup. We can leave the network running for a long time with little effort. To prevent the abuse of the compromised machines, we can let the "servers" actually be high-interaction HPs.

## 9.7 Additional Considerations

Future developments include, of course, the use of real-world tools to implement the autonomous agents (while this proof of concept could be developed in a scripting language like Python), the implementation of all possible security measures to secure and protect the agent, as well as the development of the cooperative agents postulated previously.

Moving away from a monolithic architecture (at the hypervisor level) to a swarm-like distributed architecture of agents living on every VM on the system is another valid possibility.

## 9.8 Further Reading

Guarino A., Autonomous intelligent agents in cyber offence. In: Podins K, Stinissen J, Maybaum M, editors. 5th International Conference on Cyber Conflict – Proceedings; 2013 June 4-7; Tallinn, Estonia. IEEE.

Russell S, Norvig P. Artificial intelligence: a modern approach. 3rd ed. London (UK): Pearson, 2010. ISBN: 978-0136042594.

## 10. Conclusions

Environmental factors—such as mobility, lethality, connectivity, and power constraints—limit specific functions and abilities of particular agents and the combination of these factors place an upper bound on agents' capability. Development and deployment of autonomous agents in the military context is full of caveats and will require much consideration before agents are mature enough to be used in mission-critical systems on the battlefield. Future work will require advances in multi-agent architectures, human–agent interfaces, agent reliability, agent resiliency, agent resource requirements, and agent standardization. The promises of autonomous agents are great, but much work remains to ensure they can operate in a variety of environments.

Possible data sources for autonomous agents can be largely divided into 5 categories: 1) network traffic, 2) event logs, 3) hardware sensor data, 4) OS-level software sensor data, and 5) high-level inputs. While actually collecting and labeling more of the exercise data are comparatively easy to address in exercise planning, the scale/timespan/variety of live exercises is by definition limited by resource availability. A complementary approach is to create realistic data sets on demand in emulated/simulated testbeds capable of supporting arbitrary network topologies, sufficiently real hardware, real OS/application/attack/defense software, and a combination of synthetic and real actors.

The determination of an optimal course of action may need to be computed in a small amount of time, with limited resources. An agent may be required to respond to the actions of intelligent actors (human or otherwise) with partial knowledge about the system they are protecting or the impact of its own actions. A suitable decision-making approach should take into account not only the properties of the agent itself, but the environment and context in which a decision must be made.

The agent must incorporate several capabilities. Firstly, the agent needs to perceive its environment by evaluating the state of its computer system(s), detecting changes, and deciding whether those changes are suspicious. If it does indeed detect suspicious changes, the agent must respond in an appropriate and timely manner. Secondly, the agent needs to manage trust relationships with other agents and humans through communication and cooperation. Lastly, throughout its activities, the agent needs to assess itself and its goals, and proactively initiate goal-directed actions to improve its performance and exploit opportunities when possible.

Agents will reside on devices that must operate securely in contested environments with limited connectivity. To address these cybersecurity and risk challenges, the agent should have capabilities related to ML: autonomy, cyber-risk modeling, and adversarial learning.

An agent's autonomous capability needs to satisfy the necessary and sufficient conditions for adequate situational awareness, security, and battle management by both local and remote human operators. To this end, efforts must be undertaken in systems development to preserve and optimize the cognitive resources of users. Doing that, the human user is adequately prepared to commit their own cognitive resources to the realization and consequences of the agent's action within the context of a larger plan.

## List of Symbols, Abbreviations, and Acronyms

| | |
|---|---|
| ACDA | autonomous cyber defense agent |
| AIA | autonomous intelligent agent |
| AUC | area-under-curve |
| BPF | Berkeley Packet Filter |
| C2 | command and control |
| CAS | complex adaptive system |
| CMS | Cyber Mission Forces |
| CPS | cyber-physical system |
| CPU | central processing unit |
| CRMM | Cognitive Resource Management Model |
| CyberVAN | Cyber Security Virtual Assured Network |
| DARE | Dynamic Application Rotation Environment |
| DARPA | Defense Advanced Research Projects Agency |
| DETERLab | Cyber Defense Technology Experimental Research Laboratory |
| EAPC | Euro-Atlantic Partnership Council |
| eBPF | extended BPF |
| EMCON | emissions control |
| ERM | Entity Relationship Models |
| GPS | global positioning system |
| HEAL | Human Engineering for Aeronautics Laboratory |
| HMI | human–machine interface |
| I/O | input/output |
| IDS | intrusion detection systems |
| IoBT | Internet of Battlefield Things |
| IoT | Internet of Things |
| IP | Internet Protocol |
| IPS | intrusion protection system |
| K-L | Kullback–Leibler |
| MANET | mobile ad hoc network |
| ML | machine learning |
| MORE | Multiple OS Rotational Environment |
| MTD | moving target defense |
| NATO | North Atlantic Treaty Organization |
| NSF | National Science Foundation |
| OMG | Object Management Group |
| OTE | Operational Test and Evaluation |
| OTH | over the horizon targeting |
| PCAST | President's Council of Advisors on Science and Technology |
| PfP | Partnership for Peace |
| TTPs | Tactics, Techniques and Procedures |
| UML | Unified Modeling Language |

| 1 | DEFENSE TECHNICAL |
|---|---|
| (PDF) | INFORMATION CTR |
| | DTIC OCA |

| 2 | DIR ARL |
|---|---|
| (PDF) | IMAL HRA |
| |   RECORDS MGMT |
| | RDRL DCL |
| |   TECH LIB |

| 1 | GOVT PRINTG OFC |
|---|---|
| (PDF) |   A MALHOTRA |

| 1 | ARL |
|---|---|
| (PDF) | RDRL D |
| |   K KOTT |